\documentclass[a4paper]{jpconf}

\usepackage{graphicx,color}
\usepackage{gensymb}
\usepackage{amsmath,amsfonts,amssymb}
\usepackage{hyperref}

\begin{document}
\title{Parametric study of polar configurations around binaries}

\author{C. A. Giuppone$^1$, N. Cuello$^{2,3}$}
\address{$^1$Universidad Nacional de C\'ordoba, OAC - IATE, 5000 C\'ordoba, Argentina\\
$^2$Instituto de Astrof\'isica, Pontificia Universidad Cat\'olica de Chile, Santiago, Chile.\\
$^3$N\'ucleo Milenio de Formaci\'on Planetaria (NPF), Chile}
\ead{cristian@oac.unc.edu.ar}

\begin{abstract} 
Dynamical studies suggest that most of the circumbinary discs (CBDs) should be coplanar. However, under certain initial conditions, the CBD can evolve toward polar orientation. Here we extend the parametric study of polar configurations around {detached close-in} binaries through $N$-body simulations. For polar configurations around binaries with mass ratios $q$ below $0.7$, the nominal location of the mean motion resonance (MMR) $1~:~4$ predicts the limit of stability for $e_{\rm B} > 0.1$. Alternatively, for $e_{\rm B} < 0.1$ or $q \sim 1$, the nominal location of the MMR $1~:~3$ is the closest stable region. {The presence of a} giant planet increases the region of forbidden polar configurations around low mass ratio binaries with eccentricities $e_B\sim0.4$ with respect to rocky earth-like planets. {For equal mass stars, the eccentricity excitation $\Delta e$ of polar orbits smoothly increases with decreasing distance to the binary. For $q<1$, $\Delta e$ can reach values as high as $0.4$.} Finally, we studied polar configurations around $HD~98800BaBb$ and show that the region of stability is strongly affected by the relative positions of the nodes. The most stable configurations in the system correspond to polar particles, {which are not expected to survive on longer time-scales due to the presence of the external perturber HD~$98800AaAb$.}

\end{abstract}

\section{Introduction}
\label{sec:int}

Against all odds, the number of circumbinary planets is slowly increasing. To date, 22 circumbinary {(P-type) planetary} systems have been catalogued, with 11 transiting circumbinary planets detected by \textit{Kepler} around nine binary star systems~\cite{Schwarz+2016}. The binaries eccentricities range from quasi circular orbits (\textit{Kepler}-47, $e_{\rm B}=0.02$~\cite{Orosz+2012}) to highly eccentric (\textit{Kepler}-34, $e_{\rm B}=0.52$~\cite{Welsh+2012}). All the transiting binaries have short periods ($<30$ days). Moreover, the detected planets have almost circular coplanar orbits and semi-major axis ratios with the binary $a_{\rm p}/a_{\rm B}\sim 4$. Typical planets found around binary stellar systems have a radius of the order of $10$ Earth radii and orbital periods of about $160$ days (i.e. $a_{\rm p} \sim 0.35$ au)~\cite{Martin2018}. 

Circumbinary (P-type) planets are intrinsically difficult to observe through radial velocities, which translates in a discrimination in Doppler planet searches~\cite{Eggenberger&Udry2007, Wright+2012}. Hence, these planets are more easily observed through transit methods --- provided precise considerations on the geometry of transit are considered~\cite{Martin&Triaud2014}. The occurrence rate of circumbinary planets is comparable to that of planets around single stars if the mutual inclination is always small ($< 5\degree$), while this rate may be much larger if modest inclinations ($> 5\degree$) are common~\cite{Armstrong+2014}. 
 
For circumbinary planetesimals in coplanar orbits, eccentricities evolve on a dynamical timescale, which leads to orbital crossings even in the presence of gas drag. This makes the current locations of the circumbinary \textit{Kepler} planets hostile to planetesimal accretion~\cite{Moriwaki+2004, Paardekooper+2012}. Additionally, stellar-tidal evolution models of short-period binaries show that the binary orbital period increases with time~\cite{Fleming+2018, Fleming+2019}. This translates into a larger region of dynamical instability around the binary, which could explain the lower frequency of P-type planets compared to S-type planets. 
 
Planets form within circumbinary discs, that are typically assumed to be aligned with the binary~\cite{Foucart&Lai2013,Foucart&Lai2014}. Misaligned circumbinary discs (CBDs) were considered as unlikely or at least transient; however, the observations of highly non-coplanar systems such as 99~Herculis~\cite{Kennedy+2012a}, IRS~43~\cite{Brinch+2016}, GG~Tau~\cite{Cazzoletti+2017, Aly+2018}, and HD~142527~\cite{Avenhaus+2017} suggest otherwise. More importantly, the very first confirmation of a circumbinary gas-rich disc in a polar configuration HD~98800~\cite{Kennedy+2019} motivates our study. Previous works based on Smoothed Particle Hydrodynamics (SPH) simulations studied specific conditions for polar alignment of the disc around binaries~\cite{Aly+2015, Martin&Lubow2017}. This mechanism has been further investigated providing an analytical framework to describe the polar alignment of CBDs~\cite{Zanazzi&Lai2018, Lubow&Martin2018}. {In particular, \cite{Martin&Lubow2018} conducted an extensive numerical exploration through SPH simulations.} Lastly, the symmetry breaking between prograde and retrograde CBDs reported by~\cite{Cuello+Giuppone2018,Cuello&Giuppone2019} could increase the likelihood of polar alignment.

For these reasons, it is important to further understand the stability of polar configurations around binary stars. In~\cite{Cuello&Giuppone2019}  we presented the most favourable conditions for regular movement of particles and planets around binaries. Here, we extend these results for a wider variety of configurations and analyse the case of HD~98800.

\section{Dynamics around binaries}

The secular evolution of a planet around a binary was presented in~\cite{Ziglin1975} and later investigated in~\cite{Verrier+2009}, where they reported the coupling between the inclination and the node of particles. Then, \cite{Farago&Laskar2010} studied the circumbinary elliptical restricted and general three-body problem, giving an averaged quadrupolar Hamiltonian. {In their formulation, the eccentricity of the outer particle/planet remains constant (i.e. $\Delta e=0$). This is not longer valid if the octupole expansion is considered or in full $N$-body integrations (as in \cite{Cuello&Giuppone2019} and this work).} 

An analysis of the stability of inclined \textit{massless} particles was done by~\cite{Doolin+2011} where they consider different binary eccentricities and mass ratios between its components. Furthermore, based on the study of hierarchical triple systems, several other studies investigated the circumbinary polar orbits for the restricted problem considering the octupole Hamiltonian~\cite{Ford+2000, Naoz+2013, Li+2014}. Interestingly, for near polar configurations, the inner binary can significantly excite the orbital eccentricity (up to about $0.3$ in some cases, as shown by~\cite{Li+2014}). Lastly, the evolution of polar orbits considering relativistic effects was studied by~\cite{Naoz+2017, Zanardi+2018}. The latter effects are beyond the scope of this work.

Here, we consider a binary system with total mass $M$ and individual masses $M_1$ and $M_2$, with a binary mass ratio $q=M_2/M_1$. To describe the motion of a P-type particle (planet) we use the Jacobi orbital elements semi-major axis $a$, eccentricity $e$, inclination $i$ (with respect to the binary orbital plane), mean longitude $\lambda$, longitude of pericentre $\varpi$ (alternatively argument of pericentre $\omega$), and longitude of the ascending node $\Omega$. The sub-index $B$ is used when referring to the binary orbit. Without loss of generality, we set initial conditions \{$\lambda_{\rm B}=0\degree$, $\varpi_{\rm B}=0\degree$, $\Omega_{\rm B}=0\degree$\}. The angles of a given particle are measured from the direction of the binary pericentre.

For a P-type particle the phase space has two equilibrium points at \{$\Omega=\pm90\degree, \, i=90\degree$\}, where both angles librate. For any given pair of $(\Omega,\,i)$ it is possible to calculate the separatrix F as follows \cite{Farago&Laskar2010},
\begin{equation}
\label{eq:F}
F = \sqrt{\frac{1 - e_{\rm B}^2}{1 - 5 \, e_{\rm B}^2\,\cos^2 \Omega + 4 \, e_{\rm B}^2}} \,\,\,.
\end{equation}
Then, polar alignment is expected when
\begin{equation}
\label{eq:limit}
\arcsin F < i < \pi - \arcsin F \,\,\,.
\end{equation}

Therefore, the higher the binary eccentricity $e_{\rm B}$ and the closer the node of the particle to $\Omega \sim 90\degree/270\degree$, the larger the region of polar orbits. In the next section, we numerically explore the stability for a wide range of $\{e_{\rm B},q,M_{\rm p}\}$ in polar configurations initially placed at $(i=90\degree, \Omega=90\degree)$.

\section{Stability of polar orbits}

\begin{figure}
\begin{center}
\includegraphics[height=0.32\textwidth]{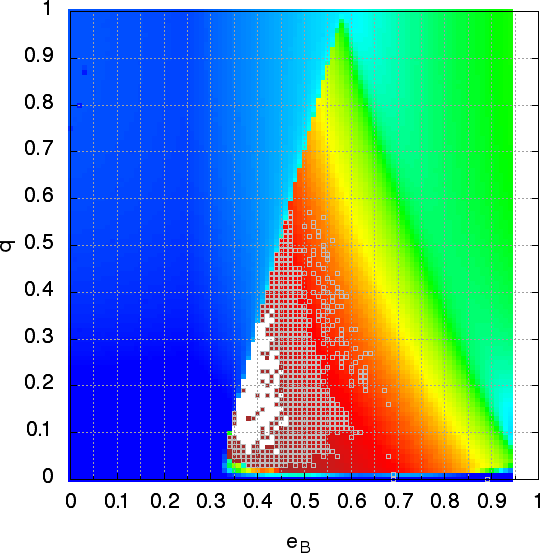}
\includegraphics[height=0.32\textwidth]{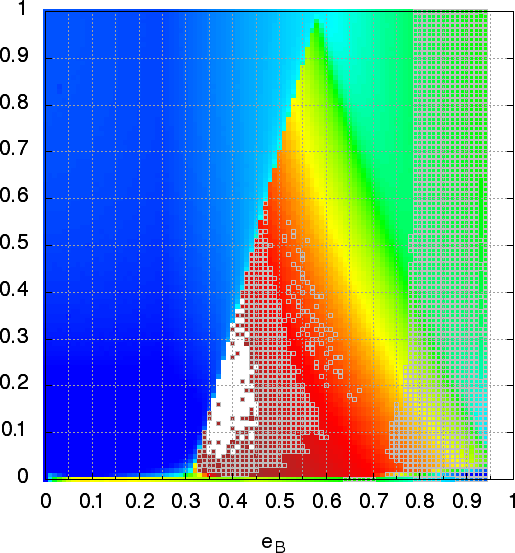}
\includegraphics[height=0.32\textwidth]{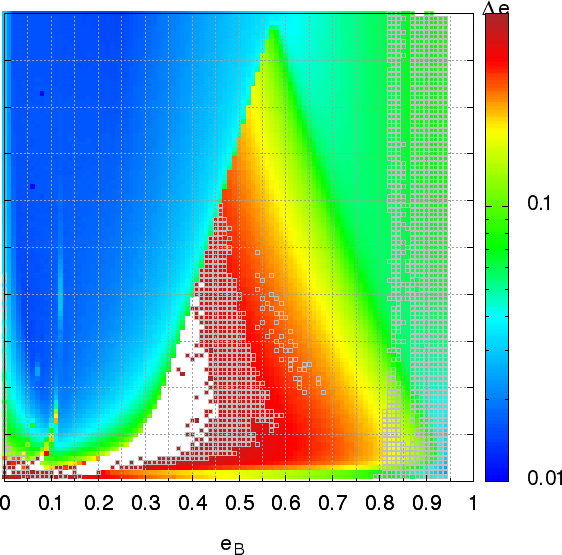}\\
\caption{Dynamical maps for polar orbits for planets with initial $a=4.5 \, a_{\rm B}$ in the ($e_{\rm B},\,q$) plane. Earth, Neptune and Jupiter-like planets are shown in the left, centre and right panels (respectively). The colour scale corresponds to $\Delta e$. The white dots correspond to unstable orbits, while the gray squares identify the chaotic orbits according to the {\sc Megno} indicator. }
\label{fig:ebJ}
\vspace{-1.5em}
\end{center}
\end{figure}

Since polar configurations are a possible outcome for the evolution of misaligned gaseous discs around sub-au binaries~\cite{Zanazzi&Lai2018,Lubow&Martin2018,Martin2018,Cuello&Giuppone2019}, it is relevant to identify which are the allowed configurations once the gas is dissipated, {in a similar way as in~\cite{Quarles+2018} for the study of the coplanar case.}

As mentioned previously, the binary sample of \textit{Kepler} systems ranges from $0.2 \leq q \leq 1$ and $0.05 \leq e_{\rm B} \leq 0.52$. Therefore, we choose to explore the stability of massless polar particles for different binary mass ratios $q=\{0.2,0.5,0.7,1\}$.
 
In all our dynamical maps each orbit was integrated for at least 365\,000 binary periods. If the particle either collides with one of the stars or escapes from the system, then it is coloured in white. In order to identify the dynamical regimes of movement, we calculated the {\sc Megno} (Mean Exponential Growth of Nearby Orbits) value $\langle Y \rangle$ for each orbit~\cite{Cincotta&Simo2000}, along with the maximum $\Delta \Omega$ and the maximum $\Delta e$ values attained during the dynamical evolution. We solve the $N$-body equations of motion numerically by means of a double-precision Bulirsch-Stoer integrator with tolerance $10^{-12}$.

In~\cite{Cuello&Giuppone2019} we studied the behaviour of planets around a binary with $e_B=0.5$, and showed that planets with masses $M_{\rm p}>10^{-5} \, M_\odot$ exhibit rapid node circulation. In this case, the coupled oscillation with the eccentricity, leads the planet to describe a pulsating sphere around the binary. In Figure~\ref{fig:ebJ}, we show the eccentricity excitation in the ($e_B,q$) plane for different planetary masses ($1 \, M_\oplus$, $17 \, M_\oplus$, $1 \, M_{\rm J}$), setting $a=4.5 \, a_{\rm B}$. {The value of $a_{\rm B}$ is similar to the observed configurations of \textit{Kepler} systems $(a_{\rm B}\sim 0.1 - 0.2$ au)}. As expected, the chaotic orbits are correlated with high values of $\Delta e$. Moreover, we observe that the larger the planetary mass, the more significant the instability region. More specifically, Earth-like planets remain with low $\Delta\Omega$ and Neptune-like planets exhibit $\Delta\Omega \sim 90\degree$; whereas the node of Jupiter-Like planets always circulates.

\begin{figure}
\begin{center}
\includegraphics[width=20pc]{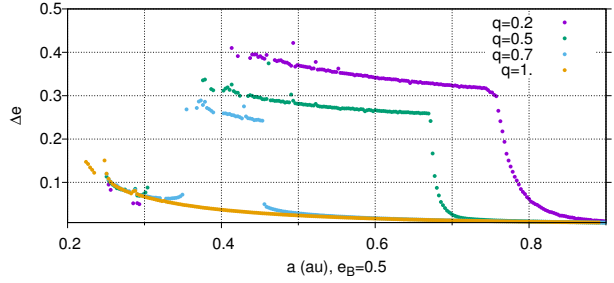}%
\vspace{-4em}
\caption{Amplitude of $\Delta e$ for different values of $q$ shown for stable orbits initially placed in a polar configuration around a binary with $e_B=0.5$. The orbits are integrated for 400\,000 orbital periods. Only the stable orbits are displayed.}
\label{fig:eb05}
\end{center}
\end{figure}

\begin{figure*}
\begin{center}
\includegraphics[width=0.49\textwidth]{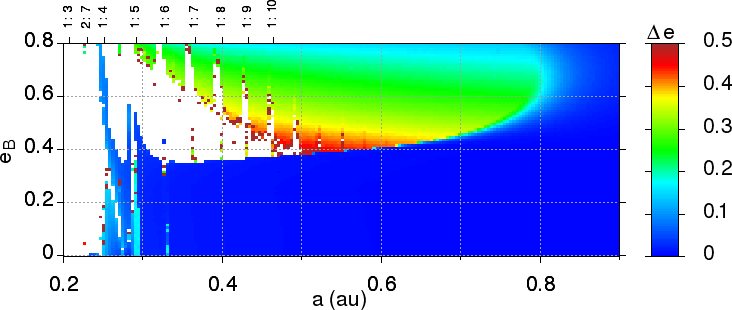} \includegraphics[width=0.49\textwidth]{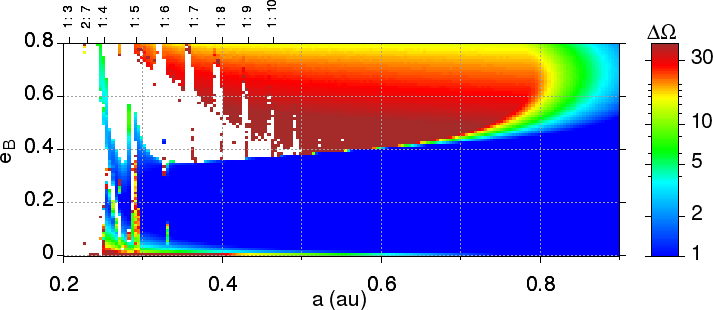} \\
\includegraphics[width=0.49\textwidth]{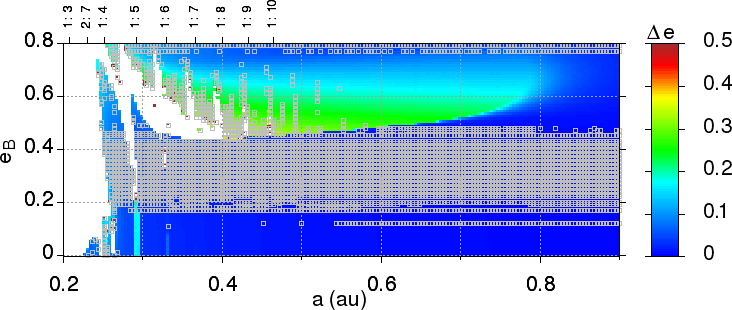} \includegraphics[width=0.49\textwidth]{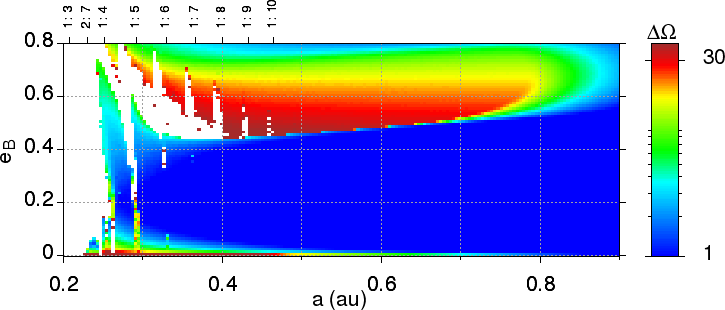} \\
\includegraphics[width=0.49\textwidth]{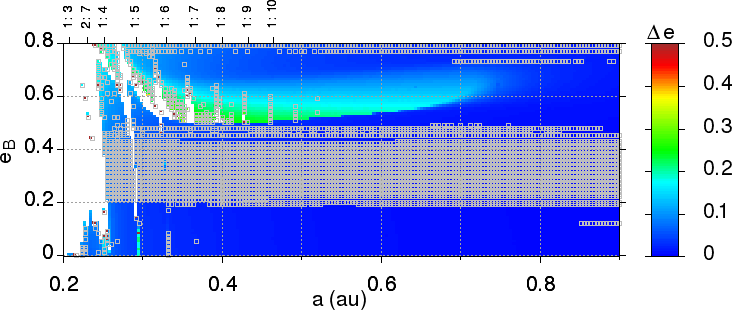} \includegraphics[width=0.49\textwidth]{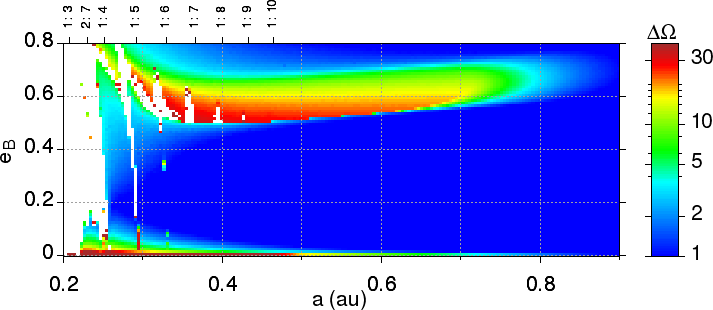} \\
\includegraphics[width=0.49\textwidth]{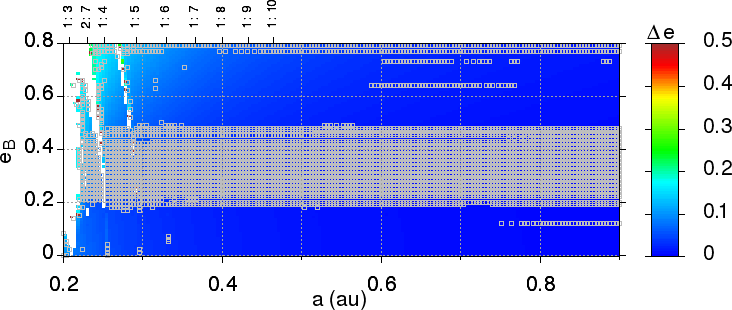}  \includegraphics[width=0.49\textwidth]{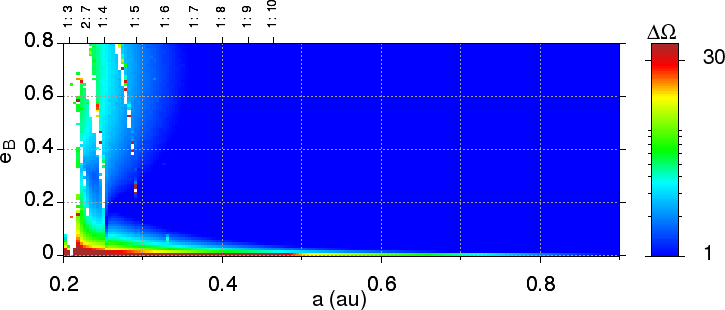} \\
\caption{Dynamical maps for polar orbits of particles around different kind of binaries {with $a_{\rm B}$=0.1~au}. From top to bottom: $q=\{0.2,0.5,0.7,1\}$. The colour scale corresponds to $\Delta e$ (left panels) and $\Delta \Omega$ (right panels); white dots are unstable orbits, while gray squares (only on the left) correspond to chaotic orbits identified with the {\sc Megno} indicator. {Initially, the particles are placed on circular orbits.}}
\label{fig:aeb}
\vspace{-1.5em}
\end{center}
\end{figure*}

\begin{figure*}
\begin{center}
\includegraphics[width=0.49\textwidth]{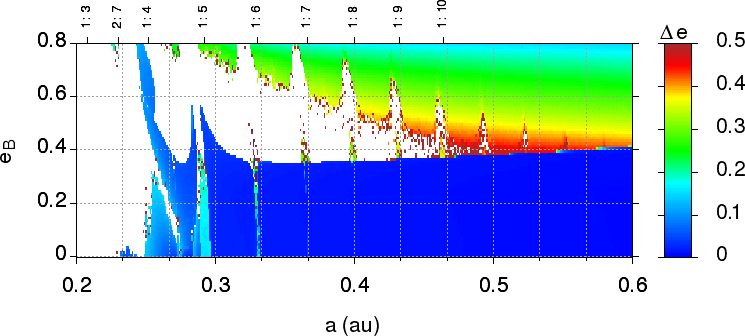} \includegraphics[width=0.49\textwidth]{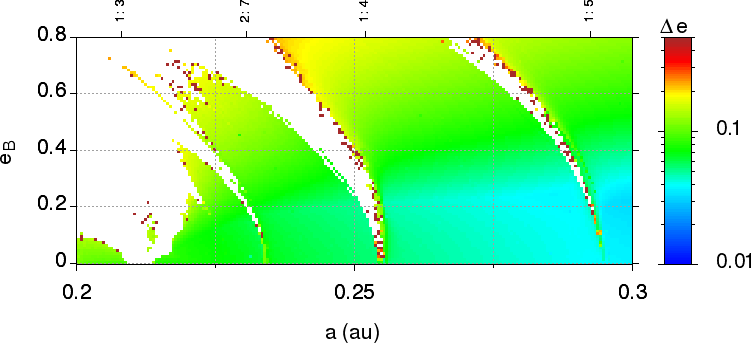} \\
\caption{Zoom of dynamical maps for polar orbits of particles around a binary $q={0.2}$ (left panel) and $q=1$ (right panel). The colour scale corresponds to $\Delta e$. The displacement of the nominal location of $1~:~N$ resonance is evident for increasing $e_{\rm B}$.}
\label{fig:aebzoom}
\vspace{-1.5em}
\end{center}
\end{figure*}

In Figure \ref{fig:eb05}, we show the eccentricity amplitude $\Delta e$ of polar orbits around a binary with $e_{\rm B}=0.5$, for different mass ratios $q$. This extends the region studied in~\cite{Cuello&Giuppone2019}. The eccentricity excitation remains below $\Delta e < 0.1$ only for binaries with $q \sim 1$. For different values of $q$, the particle experiences a significant eccentricity excitation, that strongly depends on its initial semi-major axis (contrary to the analytical model prediction in~\cite{Farago&Laskar2010}). Higher values of $\Delta e$ eventually trigger instability, which leads to the disruption of the system. {Hence, eccentricity excitation can have dramatic dynamical effects on long time-scales.}  
 
Finally, we choose to analyse the polar orbits in the plane ($a,e_{\rm B}$) for different mass ratios~$q$. In Figure~\ref{fig:aeb}, we observe (almost) vertical stripes associated to the $1~:~N$ mean motion resonances (MMRs), where eccentricity is efficiently excited. For polar orbits, the $1~:~3$, $1~:~4$, and $1~:~5$ MMRs are the most prominent features. {Their nominal location is displaced closer to the binary for increasing eccentricities $e_{\rm B}$. In Figure~\ref{fig:aebzoom}, we show a zoomed region of these maps for $q=0.2$ and $q=1$, where the limit of the stability region can be better appreciated. The latter is associated with the $1~:~N$ resonances.}

The often-quoted study of~\cite{Holman+Wiegert1999} numerically derived a stability limit for circular coplanar particles around binaries:
\begin{equation}
 a_c=a_{\rm B} \left( 1.6 + 5.1 \, e_{\rm B} - 2.22 \, e_{\rm B}^2 + 4.12 \, \mu - 4.27 \, e_{\rm B} \, \mu - 5.09 \, \mu^2 + 4.61 \, \mu^2 \, e_{\rm B}^2 \right)
\end{equation}
where $\mu=\frac{q}{1+q}$.
   
Alternatively, {we found that in polar configurations with mass ratios below} ($q<0.7$), the nominal location of the $1~:~4$ MMR predicts the limit of stability for $e_{\rm B}>0.1$; while, if $e_{\rm B} < 0.1$ or $q\sim1$, the nominal location of the $1~:~3$ MMR seems to be the innermost {limit} (see Fig.~\ref{fig:aeb} and zoom in Fig.~\ref{fig:aebzoom}). When $e_{\rm B}\gtrsim 0.5$, the perturbations on the particles become more significant {and only particles with $a \gtrsim 8 \, a_{\rm B}$ remain unaffected by the perturbation}. Then, using the MMR location, we can estimate a rule of thumb for the stability of polar orbits as:
 \begin{eqnarray}
 a_c &=& a_{\rm B} \, 4^\frac{2}{3} \, \,\,\,\,\, \mbox{ when } q\lesssim 0.5 \mbox{ and } e_B\lesssim0.4\\
 a_c &=& a_{\rm B} \, 3^\frac{2}{3} \,\,\,\,\, \mbox{ when } q\gtrsim 0.7 \mbox{ and } e_B\lesssim0.1
 \end{eqnarray}

\section{Inclined orbits around HD~98800 {\it BaBb}}

The HD~98800 system is a hierarchical quadruple stellar system, composed by two pairs of binaries
(called ‘$A$’ and ‘$B$’, or equally ‘$AaAb$’ and ‘$BaBb$’). The very eccentric binary $BaBb$ is well constrained~\cite{Boden+2005} and hosts a bright circumbinary disc discovered by~\cite{Walker+1988}. Using new data~\cite{Kennedy+2019} derived the $AB$ orbit\footnote{We note that there are small systematic differences between the supplementary material and the main text.}. The disc around ‘$BaBb$’ has an inner edge truncated by the binary $BaBb$, and the outer edge externally truncated by the companion $AaAb$. The orientation of the disc was initially thought to be coplanar with the inner binary, but higher-resolution observations suggest a different orientation recently confirmed by~\cite{Kennedy+2019}. Two solutions were proposed for the disc ~\cite{Kennedy+2019}: it is inclined either by $26\degree$ or by $154\degree$ with respect to the {sky, which corresponds to an inclination with respect to the binary\footnote{{The mutual inclination is obtained through spherical trigonometry: $i_m=\cos{i} \cos{i_{\rm B}} + \sin{i} \sin{i_{\rm B}}\cos{(\Omega-\Omega_{\rm B})}$ \cite{Giuppone+2012}}.}. of $i_m=48\degree$ or {$i_m=92.5\degree$ (respectively)}. A CBD in polar configuration is hence possible (and likely) in this system.} The millimetric dust is observed to be on circular orbits with $2.5 \text{ au} <a<4.6 \text{ au}$ . Previous $N$-body simulations for the two possible inclinations showed that the dust is ejected in less than $1$ Myr, implying that that the dust observed with {\sc alma} is embedded within a more massive gas disc that stabilises the disc against the exterior stellar perturbations of HD~98900$AaAb$~\cite{Kennedy+2019}.  

\begin{center}
\begin{table}
\caption{Dynamical configuration {for the $N$-body integrations of} HD~98800. }
\centering
\begin{tabular}{l c c c c c c}
\br
System & $a_{\rm }$ (au) & $e_{\rm }$ & $i_{\rm }$ (deg) & $\omega_{\rm }$ (deg) & $\Omega_{\rm }$ (deg) & Mass ($M_\odot$)\\
\mr
$BaBb$ & 0.982 & 0.785 & 66.8 & 109.6 & 337.6 & $Ba$=0.699, $Bb$= 0.582\\
Disc & 2.6 - 4.6 & 0.0 & 26 or 154 & 0.0	 & 17.0 & - -\\
$AB$   & 54.58 & 0.52 & 88.6 & 64. & 4.2 & $Aa$+$Ab$=1.3\\
\br
\end{tabular}\label{tab:HD98800}
\end{table}
\end{center}

\begin{figure}
 \begin{center}
 \includegraphics[width=0.47\columnwidth]{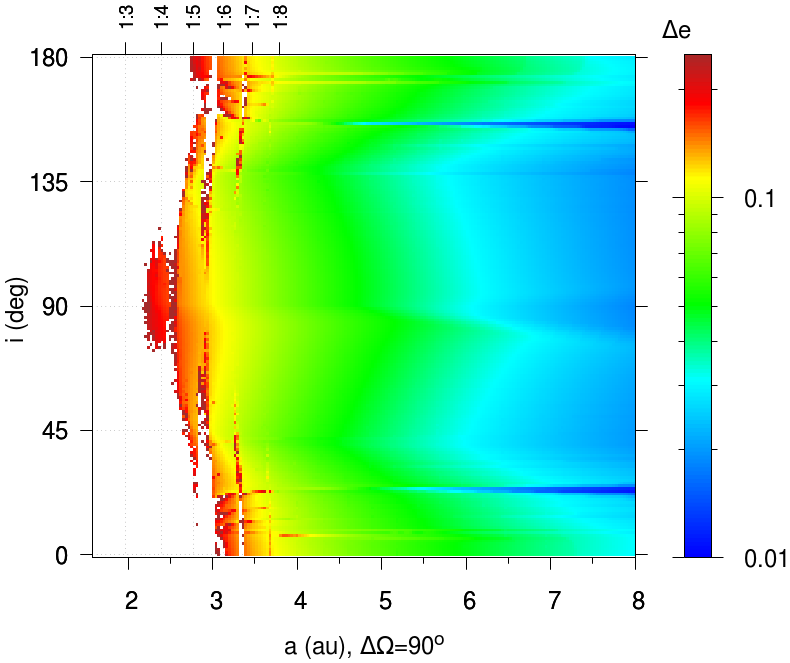}
 \includegraphics[width=0.47\columnwidth]{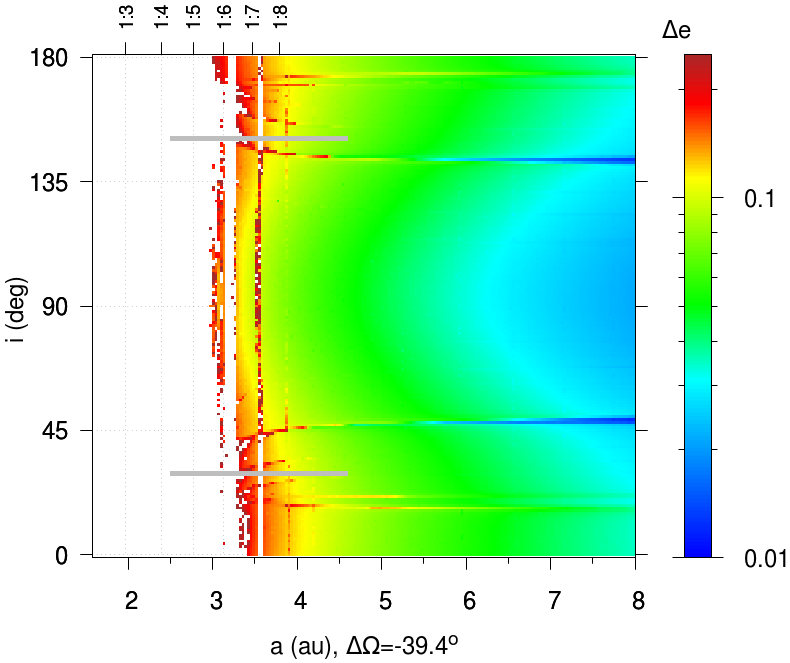}
\caption{{Dynamical maps of particles in the ($a$, $i$) plane without considering the external perturber $AaAb$. The integration is done for $3\times10^4$ orbital periods. The top labels identify the $1~:~N$ resonances.} {The fiducial system with $\Omega-\Omega_{\rm B}=90^o$ is shown in the left, while the solution published for the disc orientation ($\Omega-\Omega_{\rm B}=-39.4^o$) is shown in the right. The horizontal gray lines correspond to the two possible disc configurations, given in Table~\ref{tab:HD98800}}.}
 \label{fig:3pl}
 \vspace{-1.5em}
 \end{center}
 \end{figure}

\begin{figure}
 \begin{center}
 \includegraphics[width=0.43\columnwidth]{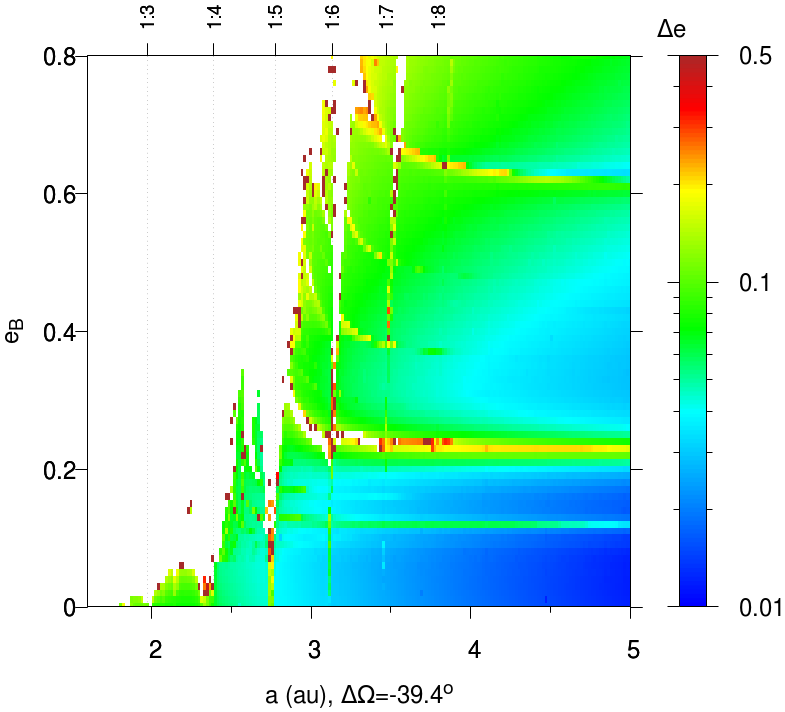}
  \caption{{Dynamical map in the ($a,e_{\rm B}$) plane for polar particles without the external perturber. The top labels identify the $1~:~N$ resonances. Stable orbits are located beyond $2$~au ($1~:~3$~MMR) for a circular binary, while for the determined $e_{\rm B}=0.785$, the region of internal stability is located beyond  $a\sim3.5$ au}.}
 \label{fig:3plaeb}
 \vspace{-1.5em}
 \end{center}
 \end{figure}

We study this system using initial conditions from Table \ref{tab:HD98800}, where the perturbing pair $AaAb$ is modelled as a single star like in~\cite{Kennedy+2019}. We note that accurate orbit determination for binaries without a complete period is extremely difficult. As a matter of fact, orbital fits exhibit high correlation between the period and the eccentricity~\cite{Giuppone+2011}, depending on the phase covered by the observations. This correlation can be observed in the table given at \verb"http://www.ctio.noao.edu/~atokovin/stars/stars.php?cat=HD&number=98800".

{In our previous section (see Fig.~\ref{fig:aeb}), the systems are initially located at the equilibrium configuration ($i=90^o$, $\Omega-\Omega_B=\pm90^o$). The binary $HD~98800BaBb$ has a mass ratio $q\sim0.83$, an eccentricity $e_B=0.785$, and $\Omega-\Omega_{\rm B}=-39.4^o$. {Then, it is meaningful to analyse the stability of polar orbits for this kind of binary.} {In Figure~\ref{fig:3pl}, we show the stability of polar particles in the ($a$, $i$) plane without considering the external perturber $AaAb$, setting $\Omega-\Omega_B=\pm90^o$ (left frame) and with the observational measured value of $\Omega-\Omega_{\rm B}=-39.4^o$ (right frame). In the latter case, the stable polar regions around the $1~:~4$ MMR disappear and only particles on polar orbits beyond the $1~:~6$ MMR are stable (that corresponds to $i=154\degree$)}}. Our integrations show that even without the external perturber, the disc of dust is unstable for $a \lesssim 3.2$.
 
{In Figure~\ref{fig:3plaeb} we show the stability of polar particles around the binary HD~98800 {\it BaBb} in the ($a$, $e_{\rm B}$) plane, without considering the external perturber and setting $\Omega=337.6^o$ and $\Omega_B=17^o$ (i.e. $\Omega-\Omega_B=-39.4^o$). When the binary is eccentric ($e_{\rm B}=0.785$), the $1~:~6$ MMR location ($a\sim3.5$ au) is the closest stable region. If the binary HD~98800 {\it BaBb} eventually circularises thought tidal interaction (see e.g. \cite{Hut1981}), then the stability of closer polar orbits would be located around the $1~:~3$ MMR ($a\sim 2$ au).}  

\begin{figure}
\begin{center}
\includegraphics[height=0.43\textwidth]{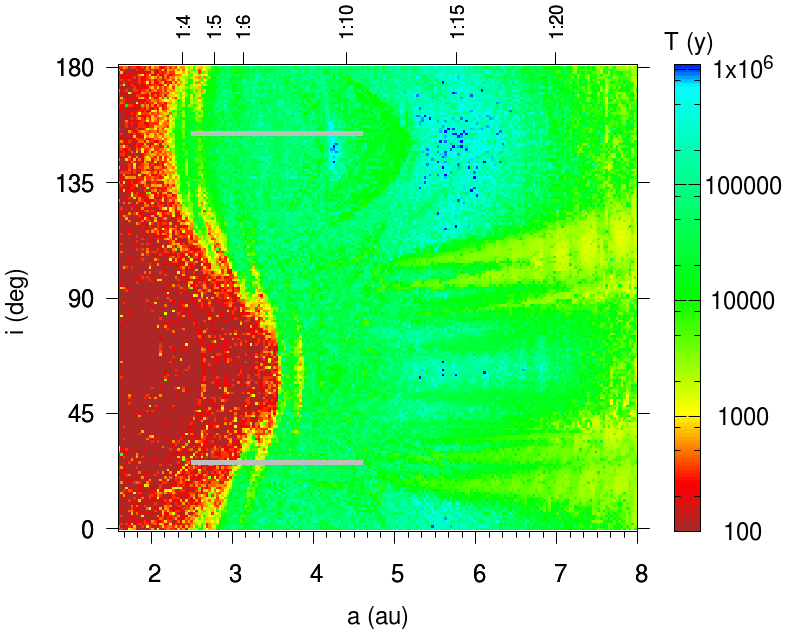}
\caption{Dynamical maps for particles around the binary $BaBb$ with external perturber. {The colour bar corresponds to the ejection time $T$}. The presence of the additional perturber $AaAb$ destroys the stability of the system, regardless of the initial inclination. Remarkably, orbits in polar configuration are more likely to survive for almost $10^6$ years. The horizontal gray lines correspond to the two possible locations for the disc, given in Table~\ref{tab:HD98800}.}
\label{fig:hd98800}
\end{center}
\end{figure}

Finally, in Fig.~\ref{fig:hd98800} we show the dynamical map in the ($a$, $i$) plane for the full four-body problem --- {coloured with the ejection time from the system} --- using the data from Table~\ref{tab:HD98800}. The most stable conditions (i.e. with survival times of $\sim 1$ Myr) are found for almost polar configuration ($i=154\degree$), for $a\sim4.2$ au and $a\sim6$ au (beyond the measured location of the disc of dust). {Interestingly, if new orbital determinations put the exterior perturber on a wider orbit, then the polar orbits around the binary $BaBb$ would become stable on longer timescales.}

\section{Conclusions}

Here we extended the previous study by \cite{Cuello&Giuppone2019} on polar configurations around eccentric binaries. The stability of single polar planets around binaries is weakly dependent on the planetary mass (see Fig.~\ref{fig:ebJ}). There is however a planetary mass threshold ($M_{\rm p} \gtrsim 10^{-5} M_\odot$) above which the planet exhibits rapid node circulation (as shown in \cite{Cuello&Giuppone2019}).

We find that the eccentricity excitation strongly depends on the distance to the binary. Then, assuming planets form in the external disc regions and then migrate inwards, this excitation eventually triggers instability. Alternatively, the eccentricity excitation remains very low for $e_{\rm B}<0.4$ when $q<0.7$. Generally, for massless particles, the orbits with $0.2<e_{\rm B}<0.4$ exhibit chaotic {\sc Megno} (see Fig. \ref{fig:eb05}).

For polar orbits around binaries with mass ratios $q$ below $0.7$, the nominal location of the $1:4$ mean motion resonance (MMR) predicts the limit of stability for $e_{\rm B}>0.1$; while if $e_{\rm B}< 0.1$ or $q\sim1$, {the nominal location of the $1~:~3$ MMR is the closest stable region to the binary (see Fig.~\ref{fig:aeb})}. 

Finally, we analysed the HD~98800 system where there is a circumbinary disc around $BaBb$ being currently perturbed by an outer binary $AaAb$. In particular, we studied the stability of the system with and without the additional binary, for any arbitrary inclination. {We find that the near polar configuration for the circumbinary disc around $BaBb$ is the more stable among the all possible disc inclinations.} We also conclude that --- in the absence of gas --- the instability would not be due to the polar configuration itself, but rather to the gravitational perturbations of the outer binary $AaAb$ (see Fig. \ref{fig:hd98800}). 

\ack{This work has been supported by research grants from CONICET and Secyt-UNC. $N$-body computations were performed at Mulatona Cluster from CCAD-UNC, which is part of SNCAD-MinCyT, Argentina. NC acknowledges financial support provided by FONDECYT grant 3170680 and from CONICYT project Basal AFB-170002. This project has received funding from the European Union's Horizon 2020 research and innovation programme under the Marie Sk\l{}odowska-Curie grant agreement No 823823.}

\section*{References}

\bibliography{polar} 
\end{document}